\documentclass[%
 reprint,
superscriptaddress,
 amsmath,amssymb,
 aps,prl,comment
showkeys 
]{revtex4-1}

\usepackage[utf8]{inputenc}
\usepackage[english]{babel}
\usepackage{color}
\usepackage{graphicx}%
\usepackage{dcolumn}%
\usepackage{bm}%
\begin{document}

\begin{acknowledgments}
{\noindent \bf  Comment on ``Inverse Square Lévy Walks \\ are not  Optimal
  Search Strategies for d $\geq$ 2''}
\end{acknowledgments}

\medskip

It is widely accepted that ``inverse square Lévy walks are optimal
search strategies because they maximize the encounter rate with
sparse, randomly distributed, replenishable targets'' \cite{ref1},
when the search restarts in the vicinity of the previously visited
target, which becomes revisitable again with high probability,
i.e., non-destructive foraging~\cite{nature1999}.  Three objecting
claims are raised in Ref.~\cite{ref1} for $d \geq 2$: (i) the capture
rate $\eta$ has linear dependence on the target density $\rho$ for all
values of the Lévy index $\alpha$; (ii) ``the gain $\eta_{\mbox{\tiny
    max}}/\eta$ achieved by varying $\alpha $ is bounded even in the
limit $\rho \to 0 $'' so that ``tuning $\alpha$ can only yield a
marginal gain''; (iii) depending on the values of the radius of
detection $a$, the restarting distance $l_c$ and the scale parameter
$s$, the optimum is realized for a range of $\alpha$.

We agree with claim (i), but as we will see, 
it is {not relevant} {in $d\geq
  2$} to whether or not inverse square Lévy walk searches are optimal
{for non-destructive foraging}.  Claim~(iii) is also correct, however
this claim was made already in
\mbox{Refs.~\cite{nature1999,prl2003,epl2004,silvia2012}}. In
particular Ref.~\cite{nature1999} showed that $\alpha=1$ is optimal
only in the {limit $l_c \to a$}, which is the main condition of
non-destructive foraging, with the quantity $l_c$ in Ref. \cite{ref1}
being none other than the distance $r_{\rm \tiny o}$ in Ref.
\cite{nature1999}.  Otherwise for large $l_c$ the optimal strategy in
the limit $\rho\to 0$ is to go along straight lines, i.e. $\alpha\to
0$.  Moreover, it is known since 2003 that a range of $\alpha$ can be
optimal (see Fig.~1 of Ref.~\cite{prl2003}, Figs.~2--3 of
Ref.~\cite{epl2004} and Figs.~1 and~S1 of Ref.~\cite{pnas2008}, none
of which are cited in Ref.~\cite{ref1}).  Crucially, \mbox{claims (i)
  and (iii)} do not {\it per se} contradict the main finding of
Ref.~[2] that $\alpha=1$ is optimal under the specific conditions of
non-destructive foraging (or of destructive foraging in patchy
landscapes)~\cite{nature1999,prl2003,epl2004,silvia2012,pre-sergey,buldyrev2001,plr,pnas2008,hide-and-seek}.

{ To test claim (ii), we have numerically simulated the identical
  model proposed in Ref.~\cite{ref1} (see Fig.~1).  The scaling for
  $\eta$ with $\rho$ proposed in Ref.~\cite{nature1999} and proved in
  Ref.~\cite{buldyrev2001} for $d=1$ does not hold in $d=2$, in
  agreement with Ref. \cite{ref1}.  However, we find, for small enough
  {\mbox{$\delta=l_c/a-1$},} that $\eta$ develops a
  maximum at $\alpha=1$ with an arbitrarily large gain relative to the
  ballistic ($\alpha\to 0$) and Brownian ($\alpha=2$) limits,
  contradicting claim (ii) about ``marginal gain'' in
  Ref.~\cite{ref1}.}

The main problem with  Ref. \cite{ref1} is that Eq.~(3) fails in the
limit $l_c \to a$ of non-destructing foraging. Eq.~(3) yields a gain
\mbox{$K_d\sim 1/[A(a^\beta-B \, l_c^\beta)] $} in Eq.~(5), with
\mbox{$\beta=-1$} for $\alpha<1$ and $\beta=\alpha-2$ for $\alpha>1$.  This
gain, which agrees with claim~(ii), is wrong in the limit $l_c \to a$.

Finally, we present a heuristic argument for the correct scaling of
$K_d$ for $d=2$ {when $l_c\to a$}.  Note that $l_c$ is
the distance at which the target stops hiding. The limit $\delta\to 0$
has biological relevance in this ``hide-and-seek''
model~\cite{hide-and-seek}.  Let $\sigma=s/a$ and
$\eta_0(\alpha,\delta,\rho,\sigma)=\eta/(\rho a)$.  When
\mbox{$\delta\to 0$}, the (radial) motion of the forager near the
border of the detection circle is essentially one dimensional, hence
the rigorous theory of the Riesz operator~\cite{buldyrev2001} on the
interval of length $L$ with absorbing ends becomes applicable.  For
$\sigma >\delta$ the efficiency increases when $\sigma$ decreases
because there are fewer large jumps leading away from the previous target that
make re-encountering it difficult.  When $\sigma\approx \delta$ the
efficiency reaches its maximum.  In the limit
\mbox{$\sigma\approx\delta\to 0$} we expect the same scaling behavior
as in \mbox{$d=1$: $\eta_0\sim \delta^{-\alpha/2}$} for $\alpha <1$
and $\eta_0\sim\delta^{-1+\alpha/2}$ for $\alpha>1$. Hence $\eta_0$
has an arbitrarily strong maximum at $\alpha=1$ when $
\sigma\approx\delta\to 0$, in agreement with Fig.~1, and in
disagreement with the title and claim (ii) of Ref.~\cite{ref1},
restoring thus the original result for non-destructive foraging in
Ref.~\cite{nature1999}  of the optimality of inverse square Lévy
flights.

\begin{figure}[t]
\includegraphics[width=1\linewidth]{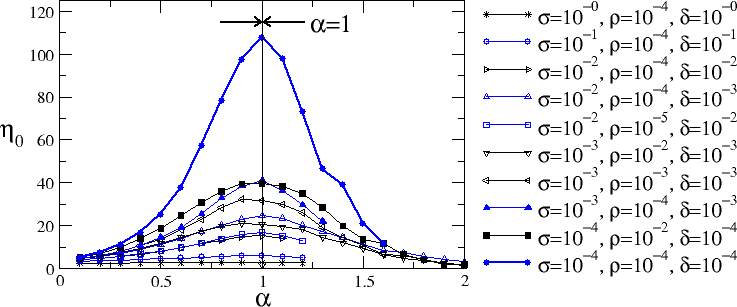}
 \caption{$\eta_0=\eta/(\rho a)$ vs. $\alpha$, for $N=10^6$ Poisson
   distributed targets on a square of size $\sqrt{N/\rho}$ with
   periodic boundary conditions, averaged over $10^5$ targets found. }
\end{figure}

\samepage{

}

\newpage

\begin{acknowledgments}

We thank H. Eugene Stanley and M. E. Wosniack for discussions and
CAPES and FACEPE for funding.  M.G.E.L., E.P.R. and G.M.V. acknowledge CNPq
grants 304532/2019-3, 305062/2017-4 and 302051/2018-0 respectively.
S.V.B. acknowledges partial support of this research by
DTRA grant No. HDTRA11910016 and through the Dr. Bernard W. Gamson
Computational Science Center at Yeshiva College.
 F.B.  acknowledges support of Grant CGL2016-78156-C2-1-R from MINECO, Spain.
G.M.V. thanks T. Macrì for pointing out Ref.~\cite{ref1}.

\bigskip \bigskip \bigskip \bigskip \bigskip

\noindent
{S. V. Buldyrev}

\vspace{0.1cm}
\hspace{0.25cm}\parbox[c]{7.0cm}{Department of Physics, Yeshiva University, New York 10033, USA}  


\bigskip
\noindent
{E. P. Raposo}%

\vspace{0.1cm}
\hspace{0.25cm}\parbox[c]{7.0cm}{Laborat\'orio de F\'{\i}sica
  Te\'orica e Computacional, Departamento de F\'{\i}sica, Universidade
  Federal de Pernambuco, Recife-PE 50670-901, Brazil; Centre d'Estudis
  Avan\c{c}ats de Blanes-CEAB-CSIC, Girona 17300, Spain; CREAF,
  Universitat Aut\`onoma de Barcelona, Cerdanyola del Vall\`es 08193,
  Spain}

\bigskip
\noindent
{F. Bartumeus}

\vspace{0.1cm}
\hspace{0.25cm}\parbox[c]{7.0cm}{Centre
  d'Estudis Avan cats de Blanes-CEAB-
CSIC, Girona 17300, Spain; CREAF,
Universitat Aut onoma de Barcelona,
Cerdanyola del Vall es 08193, Spain;
Institució Catalana de Recerca i Estudis Avançats (ICREA)}

\bigskip
\noindent 
{S. Havlin}

\vspace{0.1cm}
\hspace{0.25cm}\parbox[c]{7.0cm}{Department of Physics, Bar-llan University, Ramat-Gan 52900, Israel}

\bigskip

\noindent
    {F. R. Rusch} 

\vspace{0.1cm}
\hspace{0.25cm}\parbox[c]{7.0cm}{Departamento de Física, Universidade Federal do Paran\'a, Curitiba--PR 81531-980, Brazil}

\bigskip
\noindent
{M. G. E. da Luz}%

\vspace{0.1cm}
\hspace{0.25cm}\parbox[c]{7.0cm}{Departamento de Física, Universidade Federal do Paran\'a, Curitiba--PR 81531-980, Brazil}

\bigskip
\samepage{\noindent
{G. M. Viswanathan}%

\vspace{0.1cm}

\hspace{0.25cm}\parbox{7.0cm}{National Institute of Science and
  Technology of Complex Systems and Department of Physics, Universidade Federal
  do  Rio Grande do Norte, Natal--RN 59078-970, Brazil}
}

\end{acknowledgments}


\begin{thebibliography}{xpto}
 
 \bibitem{ref1} N. Levernier, J. Textor, O. B{\'e}nichou and
   R. Voituriez,
   Phys. Rev. Lett. \textbf{124}, 080601
   (2020).
 


   \bibitem{nature1999} G. M. Viswanathan, S. V. Buldyrev, S. Havlin,
     M. G. E. da Luz, E. P. Raposo and H. E. Stanley,
     Nature {\bf 401}, 911 (1999).


%
%
%
%



   
 \bibitem{prl2003} E. P. Raposo, S. V. Buldyrev, M. G. E. da Luz,
   M. C. Santos, H. E. Stanley and G. M. Viswanathan,
   Phys. Rev. Lett.
   \textbf{91}, 240601 (2003).
 
 \bibitem{epl2004} M. C. Santos, E. P. Raposo, G. M. Viswanathan and 
   M. G. E. da Luz,
   Europhys.
   Lett.  \textbf{67}, 734 (2004).


\bibitem{silvia2012} S. A. Sotelo-López, M. C. Santos, E. P. Raposo,
  G.~M.~Viswanathan and M. G. E. da Luz,
  Phys. Rev. E 
  {\bf 86,} 031133 (2012).


   
\bibitem{pnas2008}
  F. Bartumeus and S. A. Levin,
%
  %
 Proc. Natl. Acad. Sci. U.S.A. 
 {\bf 105}, 19072  (2008).



   


   
\bibitem{pre-sergey} S. V. Buldyrev, S. Havlin, A. Ya. Kazakov,
  M. G. E. da Luz, E. P. Raposo, H. E. Stanley and
  {G. M. Viswanathan},
  {Phys. Rev. E.} {\bf 64},
  041108 (2001). 




  
\bibitem{buldyrev2001} S. V. Buldyrev, M. Gitterman, S. Havlin,
  A. Ya. Kazakov, M. G. E. da Luz, E. P. Raposo, H. E. Stanley and
  G. M. Viswanathan,
  Physica A {\bf 302}, 148 (2001).


\bibitem{plr}
M. G. E. da Luz, E. P. Raposo and  G.M. Viswanathan, 
Phys. Life Rev. {\bf 14}, 94 (2015).
  






  
   


  
\bibitem{hide-and-seek} A. M.  Reynolds and F. Bartumeus,
  J.  Theor.  Bio.  {\bf
    260,} 98 (2009).

  

   
   
\end{thebibliography}
\end{document}